# Rapid Surface Oxidation as a Source of Surface Degradation Factor for Bi$_2$Se$_3$

Desheng Kong,[†,▼] Judy J. Cha,[†,▼] Keji Lai,[‡,§] Hailin Peng,[†, ‖] James G. Analytis,[‡, #] Stefan Meister,[†] Yulin Chen,[‡,§,#] Hai-Jun Zhang,[§] Ian R. Fisher,[‡,#] Zhi-Xun Shen,[‡,§,#] and Yi Cui[†,*]

[†]Department of Materials Science and Engineering, Stanford University, Stanford, California 94305, USA
[‡]Department of Applied Physics, Stanford University, Stanford, California 94305, USA
[§]Department of Physics, Stanford University, Stanford, California 94305, USA
[#]Stanford Institute for Materials and Energy Sciences, SLAC National Accelerator Laboratory, 2575 Sand Hill Road, Menlo Park, California 94025, USA
[‖] Current address: College of Chemistry and Molecular Engineering, Peking University, Beijing 100871, P. R. China
[▼]These authors contributed equally to this work

[*]Address correspondence to: yicui@stanford.edu.

Bi$_2$Se$_3$ is a topological insulator with metallic surface states residing in a large bulk bandgap. It is believed that Bi$_2$Se$_3$ gets additional n-type doping after exposure to atmosphere, thereby reducing the relative contribution of surface states in total conductivity. In this letter, transport measurements on Bi$_2$Se$_3$ nanoribbons provide additional evidence of such environmental doping process. Systematic surface composition analyses by X-ray photoelectron spectroscopy reveal fast formation and continuous growth of native oxide on Bi$_2$Se$_3$ under ambient conditions. In addition to n-type doping at the surface, such surface oxidation is likely the material origin of the degradation of topological surface states. Appropriate surface passivation or encapsulation may be required to probe topological surface states of Bi$_2$Se$_3$ by transport measurements.

Topological insulators (TI) represent band insulators with helically spin-polarized metallic states on the surface.[1-3] Such surface states (SS), protected by time reversal symmetry, are of fundamental research interest and attractive for applications, such as in spintronics and quantum computing. Among the candidate materials,[4-12] layer structured Bi$_2$Se$_3$ is an excellent TI with a single Dirac cone residing inside a large bandgap (~300 meV).[6, 7] SS of Bi$_2$Se$_3$ have been extensively studied by surface sensitive techniques like angle resolved photoemission spectroscopy (ARPES).[7, 13-16] Recently, several transport experiments also reveal SS: Aharonov-Bohm oscillations arising from the interference of the TI SS, observed in Sn-doped $Bi_2Se_3$ nanoribbons[17]; Shubnikov–de Haas (SdH) oscillations of the SS in bulk single crystal $Bi_2Se_3$, with pulsed, high magnetic fields up to 55T[18] and SdH oscillations of SS in gated thin flakes of $Bi_2Se_3$ exfoliated from bulk single crystals[19]. Despite these reports, many challenges still remain for studying SS by transport. For example, air exposure is shown as a n-type doping process for $Bi_2Se_3$,[20] consequently decreasing the contribution of the topological SS to transport. For pulsed, high magnetic field measurements, SdH oscillations from topological SS are smeared out after 1 to 2 hours of air exposure, a direct evidence of the degradation of surface transport properties.[18] These experimental challenges question the idealized structure and composition on the surface of $Bi_2Se_3$, which appears modified by chemical reactions under ambient conditions. In the present work, systematic transport measurements on $Bi_2Se_3$ nanoribbons, combined with surface composition analyses, provide additional evidence for the environmental doping. Based on X-ray photoelectron spectroscopy (XPS) characterizations, we show native oxide growth on $Bi_2Se_3$, both for nanoribbons and bulk crystals, which we attribute as the likely material origin of surface degradation and environmental doping. Possible mechanisms for the surface oxidation induced doping are discussed.

Transport measurements on $Bi_2Se_3$ nanoribbons provide evidence for n-type doping by air exposure. Synthesis and device fabrication of $Bi_2Se_3$ nanoribbons are reported previously,[17, 21, 22] with details given in Experimental Section. Four probe magnetoresistance, $R_{xx}$, (Figure 1a) and Hall resistance, $R_{xy}$, (Figure 1b) are measured by PPMS (Quantum Design), with a magnetic field applied perpendicular to the nanoribbon surfaces corresponding to the (0001) plane of $Bi_2Se_3$. SdH oscillations are observed both in $R_{xx}$ and $R_{xy}$. The oscillation period, $\left(\Delta(1/B)\right)^{-1}$, is determined as 86T (Supporting Information). According to the band structure of $Bi_2Se_3$, SdH oscillations of the surface electrons are expected to have a much larger period when the Fermi level is near the bottom of the conduction band,[23] so we ascribe the observed oscillations to bulk electrons. In addition, cyclotron effective mass, $m^* = 0.13\, m_e$, obtained by fitting the thermal damping of the SdH oscillation amplitude (Figure 1c; details of the analyses are provide in Supporting Information), is consistent with the value of bulk electrons from bulk single crystal measurements[20, 24]. For moderately doped samples ($n \leq \sim 5 \times 10^{18}$ cm$^{-3}$), the Fermi surface of bulk states is approximately a prolate ellipsoid in momentum space.[24, 25] Consequently, the SdH oscillation period should be proportional to $n_{3D}^{2/3}$ (Supporting Information for derivation). Figure 2 shows previous bulk crystals SdH oscillations data fitted with this relation.[23, 24, 26] By fitting the curve, we estimate the bulk electron density in the nanoribbon as $n_{3D}^{SdH} = 4.7 \times 10^{18}$ cm$^{-3}$ (Figure 2). On the other hand, effective areal electron concentration in the nanoribbon is $n_{2D}^{Hall} = 1.5 \times 10^{14}$ cm$^{-2}$, obtained from the Hall slope (Figure 1b). Hence, bulk electron concentration is $n_{3D}^{Hall} \sim 1.7 \times 10^{19}$ cm$^{-3}$, as the nanoribbon has the thickness of 88 nm, measured by atomic force microscopy (AFM, Figure 1b inset). Apparently, $n_{3D}^{Hall}$ is much higher than $n_{3D}^{SdH}$. Such large inconsistency cannot be accounted for by the possible inaccuracy of Hall measurements, which may show deviation from an actual value with a factor less than 2.[24] Most likely, a large portion of bulk electrons comes from environment doping, which exhibit low mobility and do not contribute to SdH oscillations. This effect is revealed in $Bi_2Se_3$ nanoribbons rather than bulk crystals, owing to the large surface-to-volume ratio of the nanoribbons, increasing the contribution of environmentally-doped electrons to the total conductance.

To examine the physical origin of environmental doping on $Bi_2Se_3$, surface chemistry probe X-ray photoelectron spectroscopy (XPS, SSI S-Probe XPS spectrometer with Al(Ka) source) is used to characterize the composition at the surface of $Bi_2Se_3$ nanoribbons and bulk single crystals. Si substrates with dense nanoribbon coverage are selected for XPS study, to avoid the substrate contribution in the final spectra. For as-grown nanoribbons (Figure 3a and b), a bismuth oxide peak is observed.[27] We cannot exclude the possibility that the observed surface oxide exists in the hydroxide form, which is difficult to distinguish from the oxide due to the broadness of the oxide peak overlapping the energy range of the hydroxide. We note that previous energy dispersive X-ray spectroscopy (EDS) characterizations on $Bi_2Se_3$ nanoribbons, performed in a transmission electron microscope (TEM), reveal negligible oxygen content with respect to the Bi and Se.[17, 21, 22] XPS is a surface sensitive technique, whereas TEM-EDS probes the compositions of the whole nanoribbon since the electron beam transmits through the sample. Comparison between XPS and EDS analyses indicates the oxide is formed on the surface of nanoribbons. Time evolution of the surface oxide is studied with nanoribbons stored in air (Figure 3a and b). As the air exposure time increases, $BiO_x$ peak keeps growing, corresponding to the increase of $BiO_x$ thickness. For nanoribbons exposed in air for 5 days, $SeO_x$ is also observed (Figure 3b). The absence of $SeO_x$ for samples with relatively short time exposure to air may come from the stronger tendency of $O_2$ to react with Bi, or the high vapor pressure of Se. The thickness of the surface oxide is difficult to estimate due to the random orientations of the nanoribbons. We also performed XPS study on bulk single crystal of $Bi_2Se_3$ with low carrier concentrations. These crystals are grown by slowly cooling a binary melt of elemental source powers, with excessive Se (Experimental Section). Fresh surfaces are readily available by cleaving the crystals along its natural cleavage plane, which allows us to investigate the time scale of the oxide formation. To minimize ambient exposure, the $Bi_2Se_3$ crystal is loaded into the XPS sample chamber right after cleaving (within 10s) and pumped down to vacuum with high purity $N_2$ gas flux. The $BiO_x$ signal is clearly observed in the final spectra (Figure 3c and d), which indicates the formation of the "native" oxide is very fast. The thickness of the surface oxide is estimated as ~0.38 nm, by assuming a bilayer structure with $BiO_x$ on top and $Bi_2Se_3$ underneath (Experimental Section). The oxide peak can be reduced by a short Ar ion etching for 5 minutes (10 mA, 5 kV); the weight of the $BiO_x$ peak decreases from 19% to 13% after Ar etch, further confirming the surface origin of the oxide. Spectra of the cleaved bulk crystal with longer air aging of 2 days are also acquired, revealing a dramatically increased $BiO_x$ peak. The $SeO_x$ peak is also observed in such long time aged samples, similar to the nanoribbon case. The surface oxide thickness of 2-day air exposed samples is estimated as ~ 1.94 nm (Experimental Section). Interestingly, elemental Se peak is also present in the same spectra. A tentative explanation is that some Se in the $Bi_2Se_3$ crystal grown with excessive Se powder tend to gradually precipitate out, since Se deficient state is more thermodynamically favorable under ambient condition.

The growth of native oxide (or hydroxide) on $Bi_2Se_3$ may account for many challenges in transport measurements of topological insulators. Environment doping from surface oxide growth is known for many semiconductors, such as Ge and GaN[28, 29]: Semiconductor bulk bands bend near the oxide/semiconductor interface. It may come from the induced charges in the semiconductor from the trapped charges in the oxide or at the semiconductor/oxide interface,[29, 30] or the formation of interfacial defect states giving rise to the Fermi level pinning at the interface and bulk bands bending.[28] We note that surface oxidation involving only oxygen is a p-type doping process for $Bi_2Se_3$.[16] Moisture-assisted surface oxidation of $Bi_2Se_3$ in air may exhibit distinctive interfacial defects or trapped charges different from the pure oxygen case, which results in the reversed doping effect. Another possibility is the surface "oxide" is actually a hydroxide, leading to reversed band bending. The dominant contributor to the n-type surface doping for $Bi_2Se_3$ requires further study. For continuous surface oxide growth, oxygen atoms need to diffuse into the crystal and locally distort the perfect layered crystal structure, increasing disorder on the surface. Strong surface scattering of $Bi_2Se_3$ single crystals reported by Butch et al. is thus reasonable.[23] The smearing of SS SdH oscillations of $Bi_2Se_3$ after 1-2 hours air exposure[18] may come from the environmental doping as well as the increase of surface scattering, associated with the oxidation process. Furthermore, continuous surface oxide growth is also observed in $Bi_2Te_3$,[31] another layered TI with structure close to $Bi_2Se_3$.[6, 32] Such surface oxidation appears to be a general material challenge for transport studies on this family of topological insulators. Certainly, these results implore extreme caution in handling of $Bi_2Se_3$ samples, to minimize the extraneous SS degradation effects from atmosphere exposure. Current study does not exclude alternative mechanisms contributing to the environmental doping and surface degradation process. For example, absorbed $H_2O$ serves as an electron donor for certain layered chalcogenide semiconductors, even without appreciable surface reactions.[33] The aging of $Bi_2Se_3$ surface in the atmosphere is likely the collective effect from multiple factors.

We use a simple two band model to estimate the properties of the environmentally doped layer in $Bi_2Se_3$ nanoribbons. The doped band bending region is expected as a thin layer encapsulating the nanoribbon (Schematic shown in Figure 2), due to strong screening in metallic $Bi_2Se_3$. For low field measurements, areal carrier concentration and sheet resistance of nanoribbons are approximately

$$n_{2D}^{Hall} = \left(2n_{2D}^{D}\mu_{t,2D}^{D} + n_{3D}^{SdH}t\mu_{t,3D}\right)^2 / \left(2n_{2D}^{D}\mu_{t,2D}^{D\,2} + n_{3D}^{SdH}t\mu_{t,3D}^{2}\right)$$

and $R_S = \left(2n_{2D}^{D}\mu_{t,2D}^{D}e + n_{3D}^{SdH}t\mu_{t,3D}e\right)^{-1}$, where $R_S$ is the sheet resistance, $n_{2D}^{D}$ the areal electron concentration of the environmentally doped layer, $\mu_{t,2D}^{D}$ the corresponding transport mobility, $t$ the ribbon thickness, and $\mu_{t,3D}$ the transport mobility of inner bulk electrons. As mentioned previously, $n_{2D}^{Hall} = 1.5 \times 10^{14}$ cm$^{-2}$ and $R_S$ is measured as $68\,\Omega$. From thermal damping of SdH oscillation, isotropic scattering time for inner bulk electrons $\tau_{iso,3D}$ is obtained as 0.08 ps (details in Supporting Information), which corresponds to a mobility $\mu_{iso,3D}$ of 1100 cm$^2$/Vs. For 3D electron gas, the ratio of $\mu_{t,3D}/\mu_{iso,3D}$ is larger than 1 and varies depending on microscopic scattering processes.[24, 34] For a conservative estimate, we assume $\mu_{t,3D} \sim \mu_{iso,3D}$. Therefore, we obtain $n_{2D}^{D} = 1.7 \times 10^{14}$ cm$^{-2}$ and $\mu_{t,2D}^{D} = 135$ cm$^2$/Vs. The transport mobility of the environmentally doped layer is much smaller than that of the inner bulk, which is expected as the layer will contain many interface defects due to the oxide growth.

**CONCLUSION**

In summary, the observed discrepancy between the Hall- and SdH-estimated carrier concentrations in $Bi_2Se_3$ nanoribbons indicate n-type doping by air exposure, consistent with previous measurements on bulk single crystals. XPS characterizations on $Bi_2Se_3$ nanoribbons and bulk single crystals reveal the rapid formation and continuous growth of surface oxide under ambient conditions. This oxide growth may serve as the physical origin of surface degradation and environmental doping on $Bi_2Se_3$. It presents a practical challenge for probing topological insulator surface states by transport measurements. Appropriate encapsulation or surface passivation to inhibit the surface oxidation may be required.

**EXPERIMENTAL SECTION**

**$Bi_2Se_3$ nanoribbon synthesis and device fabrication**

$Bi_2Se_3$ nanoribbons are synthesized by vapor phase growth in a horizontal tube furnace.[17, 23-24] As grown $Bi_2Se_3$ nanoribbons are grown by using 20 nm Au nanoparticles as catalyst. They are used for XPS measurements. Sn-doped $Bi_2Se_3$ nanoribbons are grown with Sn/Au (5nm/2nm) double-layer alloy catalysts. Sn-doped nanoribbons are selected for transport study. Sn effectively compensates residue bulk electrons induced from Se vacancies without significantly degrading transport mobility,[17] although Sn-doped nanoribbons remain metallic. Nanoribbons exhibit well-defined structure, with very smooth top and bottom surfaces perpendicular to (0 0 0 1) trigonal axis.

$Bi_2Se_3$ nanoribbons are mechanically transferred from the growth substrate onto a 150nm thick silicon nitride on Si substrate. Ti/Au (5nm/195nm) or Cr/Au (5nm/195nm) ohmic contacts are patterned on nanoribbons to form six-terminal configurations, through standard e-beam lithography followed by thermal evaporation of metals.

**$Bi_2Se_3$ bulk single crystal growth**

Single crystals of Bi2Se3 have been grown by slow cooling a binary melt. Elemental Bi and Se are mixed in alumina crucibles in a molar ratio of 35:65. The mixtures are sealed in quartz ampules and raised to 750 °C and cooled slowly to 550 °C, then annealed for an extended period. Crystals grown with such nominal composition ratio containing excessive Se exhibit reduced defect density and low carrier concentration.[20]

**Transport measurements**

Standard low-frequency (1,000 Hz) four-probe magnetoresistance measurements are carried out in a Quantum Design PPMS-7 instrument, Janis-9T magnet He-cryostats and a Keithley S110 Hall effect measurement system. The temperature range is 2~300 K and the magnetic field is up to ± 9 T. The field is applied to the normal direction of the nanoribbon surfaces, corresponding to the (0001) plane of $Bi_2Se_3$.

**XPS sample preparation and surface oxide thickness estimation**

XPS characterizations on nanoribbons are performed directly on as-grown substrates. High yield substrates with dense nanoribbon coverage are selected, to eliminate the contribution from the substrates to the final spectra. Flat surface of $Bi_2Se_3$ bulk single crystals is available by cleaving the crystal along its natural cleavage plane (perpendicular to its trigonal axis).

To estimate the thickness of the oxide, we assume a bilayer structure with the surface oxide on top and $Bi_2Se_3$ underneath. We obtain the atomic concentration of $Bi_2Se_3$:$BiO_x$ as 81.4 %:18.7 %, from areas under the $Bi_2Se_3$ and $BiO_x$ peak in $Bi^{4f}$ transitions. Using the electron escape depth of 1.8 nm (typical for Al Ka source), the attenuation of the $Bi_2Se_3$ peak from 100 % to 81.4 % due to the surface oxide gives the oxide thickness of ~0.38 nm for samples with minimum air exposure. For crystals aged in air for 2 days, we observe an increased $BiO_x$ peak as well as $SeO_x$ peak. Assuming again the bi-layer structure with surface oxide on top and $Bi_2Se_3$ underneath, the attenuation of the $Bi_2Se_3$ peak from 100 % to 34.1 % gives the surface oxide thickness of ~ 1.94 nm for a 2-day exposed sample.

***Acknowledgment.*** Y.C. acknowledges the supported from the Keck Foundation, and King Abdullah University of Science and Technology (KAUST) Investigator Award (No. KUS-l1-001-12). K.L. acknowledges the KAUST Postdoctoral Fellowship support No. KUS-F1-033-02. J.G.A. and I.R.F. acknowledge support from the Department of Energy, Office of Basic Energy Sciences under contract DE-AC02-76SF00515.

***Supporting Information Available.*** Details on the analyses of SdH oscillations are available free of charge via the Internet at http://pubs.acs.org.

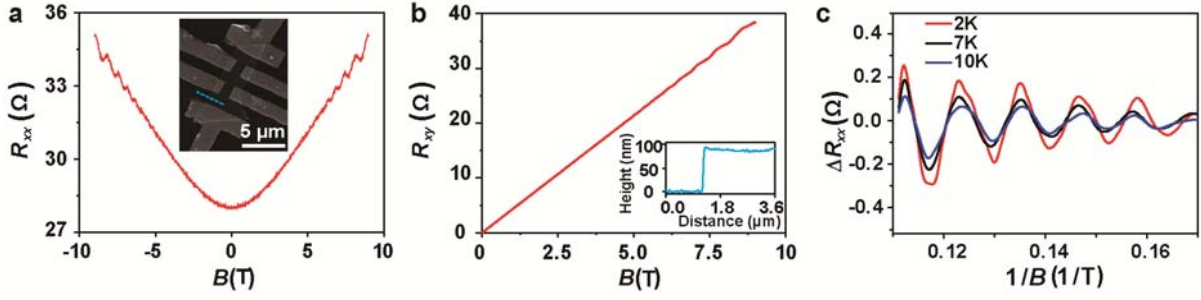

Figure 1. Transport measurements on $Bi_2Se_3$ nanoribbon device S1. (a) Magnetoresistance reveals pronounced SdH oscillations. Inset: SEM image of the device. (b) Hall resistance also shows SdH oscillations. Inset: AFM thickness profile corresponding to the line-cut of the SEM image in (a). (c) Temperature dependent SdH oscillations in magnetoresistance after background subtraction.

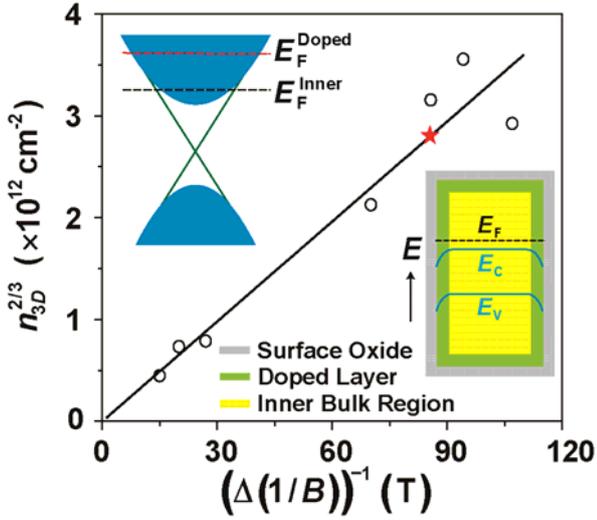

Figure 2. Plot $n_{3D}^{2/3}$ versus $\left(\Delta(1/B)\right)^{-1}$ fitted with linear relation. Dot circles are data collected from previous measurements on bulk single crystals.[23, 24, 26] Carrier concentration of the nanoribbon device is thus obtained, as marked by the red star. Upper Inset: A schematic band structure of $Bi_2Se_3$, with different Fermi levels for environmentally doped layer (red) and crystal inner region (black). Lower inset: Structure of $Bi_2Se_3$ nanoribbons consists of a surface oxide layer, heavily doped layer from environmental doping (low mobility) and moderately doped inner region (high mobility). Diagram of Fermi level, conduction and valence band edges through the nanoribbon is also shown. The existence of surface oxide is revealed by XPS characterizations (shown in Figure 3).

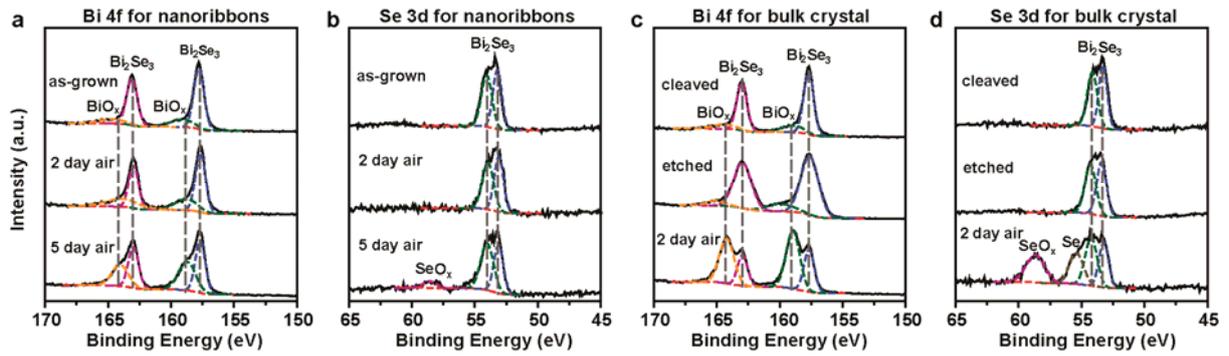

Figure 3. XPS studies on Bi$_2$Se$_3$ nanoribbons with Bi 4f (a) and Se 3d (b) spectra, including samples after synthesis, aged in air for 2 days, and aged for 5 days. XPS studies on Bi$_2$Se$_3$ single crystal with Bi 4f (c) and Se 3d (d) spectra, including samples right after cleaving (<10s air exposure), etched with Ar plasma for 5 min, and aged in air for 2 days. The observed surface oxide may exist in hydroxide form, which is difficult to distinguish from oxide due to the broadness of the oxide peak.

.